# Perspectives d'apport du projet PRIM dans le domaine du handicap

Céline Jost, Justin Debloos, Gérard Uzan
Université Paris 8
Laboratoire CHArt
Saint-Denis

Brigitte Le Pévédic
Université de Bretagne Sud
Laboratoire Lab-STICC
Vannes

Agnès Piquard-Kipffer, Caroline Barbot-Bouzit
INSHEA
Laboratoire Grhapes
Suresnes

*Abstract*— Avec le projet PRIM, nous souhaitons permettre à des personnes de créer des scénagrammes (scénarios d'interaction entre un humain et des dispositifs numériques) sans devoir se former à la programmation ou faire appel à des informaticiens. Dans PRIM, la conception du logiciel renverse les codes classiques dans le sens où il représente la pensée de l'humain (fondée sur les interactions) plutôt que la logique de l'ordinateur (fondée sur l'algorithmique) grâce à une représentation renouvelée du temps, abandonnant la *timeline* normée pour une *timeline* modifiée, propre à PRIM. Après avoir évalué l'acceptabilité et la compatibilité cognitive de cette *timeline* par 50 participants et obtenu des résultats encourageants, nous présentons, dans ce papier, les résultats de l'évaluation qualitative de l'intérêt d'un tel outil pour les participants et, plus spécifiquement, dans le domaine du handicap.

## I. Introduction

Le projet PRIM (Playing and Recording with Interactivity and Multisensoriality) a pour objectif de réunir une communauté pluridisciplinaire pour concevoir un logiciel original permettant de créer rapidement et simplement des scénagrammes, c'est-à-dire des scénarios d'interaction entre un humain et différents objets numériques [1][2]. Il répond au besoin exprimé par une grande partie de la population d'être autonome dans la création d'activités interactives. En effet, à l'heure actuelle, il est nécessaire d'utiliser des langages de programmation pour accéder aux fonctionnalités des objets connectés afin de les faire coopérer entre eux. Ce qui revient, pour cette population, à devoir choisir entre se former à la programmation ou sous-traiter à des informaticiens. Dans les deux cas, le processus de création est considérablement ralenti ou rendu complètement impossible. Pourtant, de nombreux domaines en ont besoin, par exemple l'éducation pour créer des exercices dans le cadre de la pédagogie différenciée ; la santé pour créer des activités dans le cadre de la rééducation ou des exercices de stimulation cognitive pour des patients ayant des troubles cognitifs ; l'art pour créer des œuvres numériques qui évoluent en fonction des actions du visiteur ; le théâtre pour permettre au metteur en scène de définir des interactions entre le comédien sur scène et des dispositifs numériques ; le cinéma pour créer des films 4D interactifs ; la recherche pour créer des conditions expérimentales permettant d'explorer l'impact de la technologie sur l'humain, etc.

A l'heure actuelle, il n'existe aucun logiciel permettant de créer toutes ces interactions sans programmation [3], [4]. La solution qui semble la plus simple est incarnée par l'ensemble des langages de programmation visuelles existants tels que Choregraphe pour le robot Nao [5], Blockly pour la programmation sur des objets simulés [6], [7], ou Scratch très utilisé pour apprendre la programmation à des enfants [8]. La programmation visuelle a été une révolution dans le domaine puisqu'elle s'est avérée très complémentaire à la programmation textuelle et a permis à davantage de personnes de faire de la programmation [9]. Mais elle est encore trop complexe pour permettre à tout un chacun de mettre en œuvre ses idées [10].

L'originalité du projet PRIM est de vouloir changer de paradigme en proposant un logiciel basé sur la façon de penser des humains et non sur la façon de fonctionner de l'ordinateur. Ainsi, nous souhaitons que l'humain puisse facilement créer des interactions basées sur son modèle mental, sans nécessité de les traduire dans la logique informatique, ce qui rendra le système très simple puisque perçu comme naturel. Le logiciel proposera bien un langage graphique, assimilé à de la programmation, mais basé sur un mode de réflexion propre à l'humain (ici l'interaction) plutôt que sur un mode de réflexion propre à la conception informatique et donc au langage informatique (ici l'algorithmique). L'objectif est donc de représenter la vision de l'humain et non la vision de la machine (comme c'est le cas dans les langages de programmation basés sur l'algorithmique). Dans cette perspective, la section II pose le contexte, les limites et les attendus du projet, présente les idées qui constituent le socle du logiciel à développer, puis présente le verrou scientifique majeur de ce travail. Ensuite la section III présente un prototype de logiciel développé pour proposer une solution au verrou majeur identifié et évaluer l'acceptabilité du futur logiciel afin de vérifier si le projet peut poursuivre dans sa voie. La section IV présente la méthodologie de l'évaluation menée pour valider notre proposition et vérifier si les utilisateurs s'y projettent. Ensuite la partie V présente les avis des utilisateurs sur l'utilité d'un tel logiciel. Puis la partie VI discute **ces** résultats et conclut le papier.

## II. IDÉES ET CONCEPTS

### A. Contexte : le scénagramme

Parce que notre objectif est de créer un langage différent d'un langage de programmation, plus simple, présentant une logique différente, il est exclu d'essayer de refaire ce que font déjà les langages de programmation. Ainsi, notre objectif est de créer ce que nous appelons des scénagrammes [2] et qui sont définis par « une suite d'actions effectuées par l'utilisateur et/ou par des objets numériques, alternativement, pour atteindre un but commun basé sur la stimulation cognitive ». Cela veut dire qu'un scénagramme utilise les fonctionnalités existantes des capteurs et des actionneurs présents sur les objets connectés.

### B. Idées principales

Une revue de littérature [10] a montré qu'une des raisons du succès des langages de programmation visuelle réside dans la conception même de leur interface qui est composée de plusieurs zones clairement identifiées et qui aident à la programmation : zone où se trouvent les éléments de programmation (composants), zone où l'on construit le programme, zone de paramétrage des composants et zone d'exécution du programme. Nous retrouvons exactement les mêmes zones dans les éditeurs de montage vidéo et de conception musicale qui sont des logiciels facilement utilisés et qui permettent de faire de la création.

L'autre succès de la programmation visuelle vient de sa manipulation facile avec des composants que l'on déplace d'une zone à l'autre et qui graphiquement donnent des indices pour aider à la programmation par leur forme et/ou par leur couleur. C'est également une caractéristique que l'on retrouve dans les deux autres types de logiciels susmentionnés.

Ainsi, le projet PRIM cherche à s'inspirer des langages de programmation visuelle mais également des éditeurs de montage vidéo [11]–[16] et de conception musicale [17] qui possèdent les mêmes points forts mais qui présentent une approche différente pour permettre la création par l'utilisateur.

### C. Verrou scientifique : le temps

La plus grande différence que l'on observe entre les langages de programmation visuelle et les deux autres types de logiciels réside dans la représentation et la gestion du temps. En effet, les premiers sont basés sur un temps relatif et événementiel où chaque action arrive après la précédente mais à un temps indéfini. Certaines actions peuvent se passer à n'importe quel instant et d'autres peuvent même ne jamais arriver. Cette incertitude, omniprésente lorsqu'on interagit avec un humain, est totalement absente des seconds types de logiciels qui sont basés sur un temps réel. Dans ce cas, chaque action arrive à un instant T précis et à une durée précise. Le temps se poursuit inlassablement sans jamais s'arrêter et il est impossible de prendre en compte des actions incertaines. Ces deux temporalités, qui sont incompatibles par nature, existent de façon séparée (soit dans des logiciels différents, soit dans des zones différentes d'un même logiciel comme pour Chorégraphe, par exemple). Mais il n'existe aucun logiciel où elles co-existent en utilisant la même charte graphique. Dans la majorité des cas, le temps est représenté par une ligne de temps mais dans le cas des langages de programmation visuelle, c'est la construction du programme qui construit, au fur et à mesure, une ligne de temps (horizontale ou verticale), tandis que dans le cas des éditeurs de montage vidéo ou de conception musicale, la ligne de temps existe et est représentée par une *timeline* sur laquelle l'utilisateur construit sa vidéo ou sa musique. Dans le premier cas, la ligne de temps est donc plutôt implicite et l'utilisateur doit la reconstruire mentalement, alors que dans le second cas, elle est explicite et semble représenter directement le modèle mental de l'utilisateur.

Dans le cadre du projet PRIM, nous faisons l'hypothèse que les seconds types de logiciels sont plus simples à utiliser parce qu'ils se basent sur une *timeline*. D'une part, elle est facile à utiliser et à manipuler, et d'autre part elle pourrait éviter la complexité cognitive de traduire mentalement la pensée humaine vers la pensée informatique et inversement.

Cette proposition possède un verrou scientifique majeur. Les utilisateurs sont-ils capables de s'habituer et d'accepter une *timeline* qui ressemble à celles des éditeurs de montage et de conception mais qui manipule un temps événementiel ? En effet, son utilisation risque de demander un effort cognitif différent, et de déconstruire les habitudes prises avec une *timeline* basée sur le temps absolu.

## III. SCENAPROD : PREMIER PROTOTYPE

ScenaProd (pour « Production de scénagrammes ») est un prototype réalisé dans le but de mettre en œuvre une *timeline* événementielle afin de répondre à la problématique présentée dans la section II.C. La Fig. 1 montre une présentation générale du prototype qui contient un menu pour créer, jouer ou stopper un scénagramme ainsi que trois zones facilement identifiables (palette des composants, paramétrage, édition). Les utilisateurs peuvent utiliser 4 composants rappelant des usages classiques pour faciliter leur imagination et leur projection dans un futur logiciel complet et sont suffisamment nombreux pour utiliser la *timeline*. L'utilisateur peut jouer du son, afficher du texte, afficher une image ou attendre que quelqu'un appuie sur une touche du clavier. Lors de la lecture du scénagramme, les textes et les images s'affichent dans une petite fenêtre qui permet de visualiser le scénagramme en cours de lecture.

Au niveau de la *timeline*, le prototype propose un trait en pointillé pour représenter un temps discontinu et se comporte comme une partition de musique, en passant à la ligne afin d'éviter un défilement infini horizontal qui est plus difficile à manipuler que le défilement vertical. A chaque fois qu'un utilisateur place un composant sur la *timeline*, une croix noire apparaît à la suite. C'est un menu contextuel qui permet de faire des altérations de la *timeline*. Par exemple dans ce prototype, il est possible de faire une duplication de la *timeline* de façon à obtenir deux lignes de temps, en parallèle et autonomes. C'est ce point particulier qui est l'objet de notre évaluation. En effet, des *timelines* dupliquées (visibles sur la figure par des pointillés verticaux qui les relient) sont autonomes et, en cours de lecture, il n'existe aucune synchronisation temporelle entre les deux. Or, dans les logiciels de montage vidéo, il existe une synchronisation temporelle verticale : tout élément placé sur un même axe vertical est affiché ou lu au même instant. C'est différent dans le cas de ScenaProd où des éléments placés sur le même axe vertical peuvent être lus à des moments différents comme sur une partition de musique.

C'est ce point précis qui constitue un verrou scientifique majeur. L'utilisateur peut-il accepter cette désynchronisation inhabituelle et contraire à ses habitudes ? Et peut-il se projeter dans un tel logiciel ?

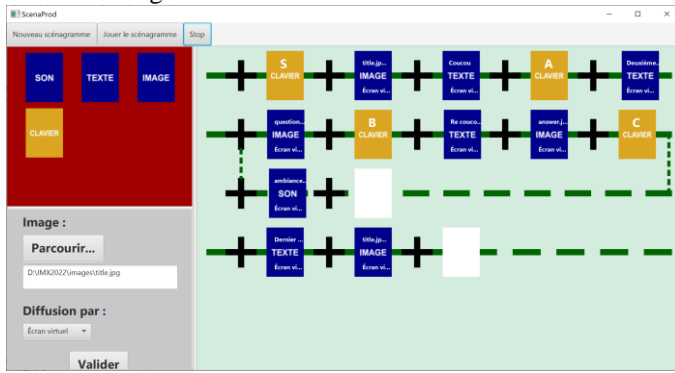

Fig. 1. Capture d'écran de ScenaProd

## IV. ÉVALUATION

L'objectif de l'évaluation était de vérifier l'acceptabilité de la *timeline* et la projection des utilisateurs dans ce type de logiciel. Chaque passation se déroulait à distance sur la plateforme Zoom et durait au maximum 30 minutes. Les critères d'inclusion étaient donc l'accès à un ordinateur et à une connexion Internet. Pour effectuer l'évaluation, les participants prenaient le contrôle de l'ordinateur de l'expérimentateur (sous MacOs) qui restait présent tout au long de la séance afin de donner des instructions, répondre aux questions ou aider les participants, si nécessaire. L'évaluation était divisée en trois étapes : (1) Les participants (ou les parents dans le cas des mineurs) devaient signer un formulaire de consentement et étaient informés qu'il n'y avait aucun enregistrement, que les données seraient anonymisées et qu'ils pouvaient arrêter n'importe quand. (2) Les participants devaient suivre les instructions de l'expérimentateur sous forme d'un tutoriel pour les amener à créer trois scénagrammes de difficulté croissante. Les tâches étaient choisies de manière à ce que chaque participant vive la même expérience de lecture de scénagramme qui était notre sujet d'étude. Le troisième scénagramme à construire avait pour but d'exposer le participant à différents stimuli simultanés qui étaient positionnés sur des axes verticaux différents. L'objectif de l'expérimentateur était alors de s'assurer que chaque participant avait bien vu et expérimenté la lecture de ce scénagramme complexe[1]. (3) Les participants devaient remplir un questionnaire. La première partie du questionnaire était le F-SUS (la version française du System Usability Scale) [18] qui nous permettait de vérifier si l'interface graphique était facile à utiliser sans constituer un biais ou un frein pour l'évaluation de la *timeline*. La deuxième partie, composée de 10 questions utilisant la même échelle que le SUS (échelle de Likert en 5 points), posait des questions plus spécifiques sur le logiciel, la *timeline* et les projections du participant (voir TABLE I. ). La troisième partie comportait les informations signalétiques du participant ainsi que ses opinions sur différents sujets.

En plus de cela, l'expérimentateur devait noter la durée totale de la session, le nombre de questions posées, le nombre de fois où les participants avaient été bloqués et les commentaires des participants s'il y en avait.

Avant cette évaluation, nous avions effectué une évaluation préliminaire avec 5 participants pour tester le protocole et nous assurer qu'une session durait moins de 30 minutes.

TABLE I. QUESTIONS POSÉES AUX PARTICIPANTS

| |
|---|
| Q1. J'ai éprouvé des difficultés à effectuer la tâche demandée. |
| Q2. Je trouve que ScenaProd ressemble à un logiciel de montage vidéo. |
| Q3. Je pense qu'il est nécessaire d'avoir des compétences en informatique pour utiliser ScenaProd. |
| Q4. Je trouve qu'il est difficile de comprendre comment positionner les blocs sur la ligne de temps. |
| Q5. Je pense que ScenaProd pourrait être utile dans le cadre de mon activité. |
| Q6. Lorsqu'on clique sur « Jouer le scénagramme », je pense que l'avancement dans le scénagramme est facile à comprendre visuellement. |
| Q7. Je trouve que la gestion du temps est déstabilisante. |
| Q8. Je trouve qu'il est difficile de comprendre que chaque ligne de temps possède son propre temps. |
| Q9. Je trouve qu'il est facile de faire une duplication de la ligne du temps. |
| Q10. Je pense que je suis capable de faire des nouveaux scénagrammes sans aide. |

## V. RÉSULTATS

### A. Participants

L'étude a été menée avec 50 participants venant de différentes régions françaises (31 F, 19 H), âgés de 12 à 75 ans (moyenne : 34,5 ans ; écart-type : 15,4). Les hommes étaient âgés de 12 à 52 ans (moyenne : 28,9 ans ; écart-type : 12,3) et les femmes de 12 à 75 ans (moyenne : 38 ans ; écart-type : 15,8).

La majorité des participants étaient dans la vie active (33). Les autres étaient collégiens (4), lycéens (2), étudiants (9) ou retraités (2). Parmi les participants dans la vie active, 17 avaient une profession intermédiaire, 12 étaient cadres ou dans des professions intellectuelles, 2 étaient employés, 1 était artisan et 1 était en reconversion professionnelle.

Parmi les participants, 11 avaient une activité en lien avec la santé, 12 avec l'informatique et 5 faisaient de l'enseignement. Trente participants avaient déjà utilisé un logiciel de montage vidéo. Et à l'issue de l'évaluation, 41 personnes ne connaissaient aucun outil similaire à ScenaProd. Les 9 autres avaient cité des logiciels de didacticiel, Microsoft Powerpoint ou des logiciels de montage vidéo.

Chaque participant a manipulé le logiciel en moyenne 18 minutes par session (médiane : 17,5 ; écart-type : 5,1 : min : 9 ; max : 33). Cinq participants ont eu besoin de l'aide de l'expérimentateur pour avancer (8 interventions au total), tandis que 18 participants ont posé des questions, souvent pour obtenir une validation de l'action à faire, pour un total de 47 questions.

---

[1] Voir ici : https://youtu.be/j8xpcJCUEXc

## B. Acceptabilité de la timeline

Le prototype a obtenu un score de F-SUS de 84 avec un écart-type de 8,12 indiquant une acceptabilité quasiment excellente, ce qui indique que le prototype a permis de mettre en œuvre et d'évaluer la timeline dans les meilleures conditions possibles (l'évaluation du prototype est hors de propos dans ce travail). Ce qui est intéressant c'est que 18% des participants ne connaissaient aucun outil similaire mais que 60% reconnaissaient l'inspiration venant des outils existants.

La Fig. 2 montre les résultats aux dix questions présentées dans la TABLE I. Les résultats sont très prometteurs puisque 92% des participants se sentent capables de refaire un scénagramme tout seul (Q10), 86% ont trouvé que la gestion du temps n'était pas déstabilisante (Q7) et 80% ont trouvé qu'il n'était pas difficile de comprendre que chaque ligne possède son propre temps (Q8) sachant que tous les participants ont bien compris la gestion du temps à l'issue de l'évaluation.

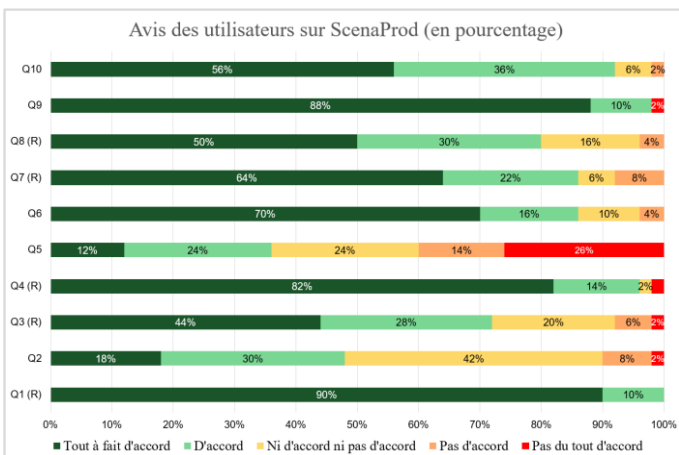

Fig. 2. Réponses aux questions de recueil d'avis des utilisateurs. Les « R » indiquent les questions pour lesquelles le codage a été inversé. Il faut donc interpréter les résultats comme si la question était positive.

En plus de ces questions, les résultats des questions d'opinion indiquent que 96% des participants ont compris que la timeline représente la progression du scénagramme et non le temps réel comme dans une vidéo ; 98% pensent qu'il est facile de comprendre et de s'habituer à cette gestion du temps ; 84% n'ont pas été dérangés par la non synchronisation verticale du temps.

Cette évaluation a permis de montrer que la synchronisation verticale peut disparaître sans perturber l'utilisateur afin de proposer une ligne de temps relatif/événementiel comme base de notre futur logiciel.

La suite de ce papier s'intéresse au ressenti des participants sur l'utilité de l'outil dans le futur et dans le cadre de leur activité professionnelle.

## C. Utilité de ScenaProd

En premier lieu, il a été demandé aux participants de donner leur avis sur l'utilité de ScenaProd lorsqu'il sera complet. Au total, 48 participants ont donné leur avis pour un total de 546 mots qui ont donné 78 propositions, les deux autres participants indiquant ne pas avoir d'avis. Ces 78 propositions ont pu être classées en 5 catégories : 17 propositions en lien avec la création, 11 perspectives dans le domaine du handicap, 20 avis de support pour l'apprentissage, 18 avis de support pour la communication et 12 propositions diverses. Cette dernière catégorie contient deux propositions isolées donc inclassables ainsi que 10 propositions qui semblent émerger directement des consignes de l'évaluation et donc possiblement être des réponses influencées (7 concernant le montage vidéo et 3 concernant la création de scénagrammes).

## D. Exemples d'utilisation dans l'activité professionnelle

En second lieu, il a été demandé aux participants de donner, s'ils le pouvaient, des exemples d'utilisation de ScenaProd dans un contexte qui est propre à leur activité. Au total 41 participants ont donné leur avis pour un total de 456 mots qui ont donné 55 propositions. Parmi les 9 autres participants, 3 ont indiqué qu'ils voyaient des exemples sans donner de précision et 6 ont indiqué ne pas avoir d'exemples à donner. Les 55 propositions ont pu être classées en 4 catégories, similaires à la question précédente : 9 propositions en lien avec la création, 10 perspectives dans le domaine du handicap, 14 avis de support pour l'apprentissage et 19 avis de support pour la communication. La Fig. 3 montre le total des propositions faites par catégorie et en fonction de l'utilité à long terme de ScenaProd et des exemples d'utilisation.

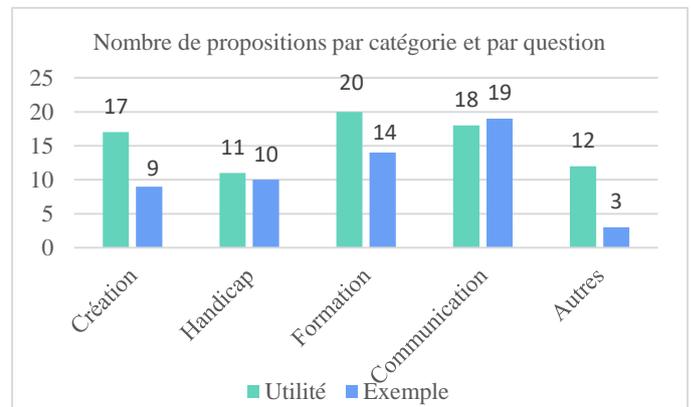

Fig. 3. Nombre de propositions par catégorie et par question

## E. Commentaires supplémentaires

Enfin, il a été demandé à chaque participant si elle ou il avait des commentaires supplémentaires à faire. La réponse a été négative pour 33 participants, et les 17 autres ont fait des commentaires pour un total de 237 mots : 8 participants ont fait des compliments sur ScenaProd, 3 ont indiqué leur curiosité de voir l'outil final, 2 ont indiqué avoir besoin de plus de temps pour pouvoir se prononcer, 2 ont fait des remarques d'ordre ergonomique et 2 ont fait des remarques d'ordre pratique.

## VI. DISCUSSION ET CONCLUSION

### A. Résultats principaux

Les résultats montrent que les participants se sont largement projetés dans l'utilisation de ScenaProd. En effet, 96% des participants ont trouvé une ou plusieurs utilités à ScenaProd lorsqu'il sera complet et 88% des participants pensent qu'il serait utile dans le cadre de leur activité professionnelle. C'est très positif car cela confirme que le prototype proposé, bien que très simple, est suffisamment complet pour permettre aux utilisateurs de se projeter dans une utilisation future. Et cela confirme également que l'outil est assez simple pour permettre une utilisation par tous, rapidement. Spontanément, et après seulement 18 minutes de manipulation en moyenne, les participants pensent à 4 champs de compétences pour ScenaProd à travers 118 avis. Sur ces 118 avis, 32,2% évoquent notre prototype comme moyen de communication ou de présentations, 28,8% comme outil de formation et d'apprentissage, 21,2% comme logiciel de création et enfin 17,8% comme solution à des problèmes dans le domaine du handicap.

*B. Communication et présentations*

Une majorité des propositions des participants évoquent la possibilité d'utiliser ScenaProd pour faire des exposés ou des présentations (60%) et voient notre prototype comme une alternative pour faire des diaporamas (18%). Le reste des propositions est moins unanime : 3 personnes pensent que ScenaProd peut permettre de communiquer sur les réseaux sociaux, 5 personnes qu'il peut représenter une interface avec un robot, compléter une information verbale, être un outil de présentation adapté, proposer des ressources interactives ou, enfin, permettre de créer des albums parlants pour la famille.

*C. Formation et apprentissage*

Il est intéressant de constater que dans les propositions classées dans cette catégorie, les participants ont pensé à trois usages. En premier les participants ont évoqué ScenaProd comme outil d'assistance de l'enseignant lors de la formation pour des publics variés (44,1%). Mais ils ont également vu ScenaProd comme plateforme d'auto-apprentissage donc à utiliser de façon autonome (38,2%). Enfin, plusieurs participants ont vu, dans notre outil, le potentiel de stimuler les utilisateurs, soit par sa nature multisensorielle, soit par sa nature ludique (17,7%).

*D. Création*

Les processus de création sont bien représentés dans les résultats. Il est intéressant de constater, au vu des inspirations de ScenaProd et de son objectif (qui est la création), que 26,9% des commentaires ont évoqué la possibilité de faire du montage vidéo, photo ou d'animation et 26,9% de la programmation (classique, de machines industriels ou de domotique). Mais d'autres disciplines ont également été citées et cela montre que ScenaProd peut favoriser le processus de création : 26,9% des commentaires ont évoqué les histoires interactives, 11,6% les créations artistiques et 7,7% les jeux vidéo.

*E. Handicap*

Cette dernière catégorie est assez transversale car, même s'il y a 21 commentaires spécifiques à la question du handicap, on trouve en réalité des propositions en lien avec le handicap dans les trois autres catégories. En faisant des regroupements et en supprimant les doublons, on peut répertorier 4 catégories : bénéficiaire, discipline, évolution de la personne, outil d'assistance.

Concernant les bénéficiaires, les participants ont mis en évidence l'utilité de ScenaProd pour les personnes autistes, pour les personnes âgées et pour les enfants. En plus d'être un outil de compensation générale pour le trouble du spectre autistique, ScenaProd pourrait permettre de communiquer à l'aide de pictogrammes et de sons et de visualiser les rituels de la journée. Pour les personnes âgées, il permettrait de mettre en place des scénarios d'assistance. Et pour les enfants il serait un bon outil pour leur éveil et pour stimuler leur créativité.

Concernant les disciplines, les participants en ont répertorié trois : ergothérapie, informatique, domotique. Notre prototype permettrait de créer des scénarios d'activités, d'occupations ou de tâches à réaliser, ce qui est le cœur de l'approche des ergothérapeutes. Il est également vu comme un outil capable de former à l'informatique ou qui est capable de donner la possibilité à des personnes en situation de handicap de comprendre et de réaliser des petits programmes. Enfin, il est également vu comme un logiciel permettant de contrôler facilement un système domotique ou des objets connectés.

Concernant l'évolution de la personne, les participants ont vu en ScenaProd un potentiel pour proposer des apprentissages adaptés, de la rééducation mnésique ou des réapprentissages, de la stimulation ou de l'entraînement (cognitif, sensoriel, créatif) et du suivi de soin. Concernant les outils d'assistance, les participants voient ScenaProd comme un facilitateur de communication. D'une part, il peut permettre à la personne en situation de handicap de communiquer (à l'aide de différentes stratégies), mais il peut, d'autre part, permettre aux autres de communiquer via des visites virtuelles, des guides personnalisés, des systèmes d'aide à domicile.

*F. Perspectives*

Les résultats montrent un très fort potentiel de ScenaProd pour stimuler la créativité et l'apprentissage et pour aider les personnes en situation de handicap. La suite du projet PRIM va donc explorer son utilité dans le domaine de l'éducation spécialisée où les besoins sont manifestes. Depuis plusieurs années, les robots sont de plus en plus utilisés dans ce domaine. Mais les robots montrent des limites de fiabilité et les enseignants ont souvent besoin de faire des modifications dans les scénarios pour adapter la situation d'apprentissage afin de combler les lacunes du robot. Ces modifications peuvent être longues ou impossibles et doivent être anticipées dans le calendrier de remédiations éducatives [19]. Dans ce contexte, un logiciel qui permette à des personnes de modéliser et matérialiser des interactions entre une personne et un système informatique sans avoir besoin de se former à la programmation aurait toute sa place. L'enseignant ou le rééducateur pourraient alors plus facilement adapter les actions des objets programmés aux besoins de l'enfant. Ce dernier pourrait actionner des contacteurs où les actions déclenchées mettraient en jeu de la multimodalité sensorielle. Ainsi, pour un enfant sourd, le toucher d'une photo avec un plat cuisiné déclencherait l'oralisation du nom du plat ou le geste associé en LSF (vidéo), associé à un souffle d'air chaud (plat chaud) ou froid (par exemple une glace). Ces mêmes associations pourraient avoir lieu dans le domaine des émotions. La vidéo d'un visage ou d'une attitude corporelle pourrait être associée à des sons ou des musiques appropriées.

*G. Conclusion*

Notre prototype ScenaProd, bien que très incomplet, a reçu une forte adhésion des 50 participants à l'évaluation. Les 4 catégories qui ont émergé montrent que notre proposition atteint ses objectifs. Premièrement, ScenaProd est vu comme un outil de communication similaire au logiciel PowerPoint, ce qui est une comparaison très positive d'autant que le parallèle portait sur la simplicité d'utilisation et que cette simplicité était l'un des premiers critères de conception de notre interface, un second étant la prise en compte de la multisensorialité au-delà du texte-image et de l'audio-visuel standard. Deuxièmement, il est vu comme un outil adapté à l'apprentissage, ce qui est le but même des scénagrammes, donc notre objectif est atteint pour cette version. En effet, produire un scénagramme signifie programmer des interactions entre un humain et des dispositifs numériques dans un but commun basé sur la cognition. Troisièmement, notre prototype a inspiré des activités de création, ce qui est également un de nos objectifs et qui montre que les inspirations de notre interface se retrouvent dans les ressentis des participants. Notre prototype réussit, pour le moment, à être un hybride des langages de programmation visuelle et des éditeurs de montage vidéo et de conception musicale. Quatrièmement, ScenaProd a largement été cité comme outil d'aide aux personnes en situation de handicap, ce qui est très encourageant pour apporter une aide supplémentaire à ces personnes.

Cependant, il faut nuancer ces résultats car, malgré l'attention portée à l'écriture des consignes et à l'animation, certains biais (pré-suggestion, relances, etc.) dû aux consignes ou à l'exemple de scénagrammes peuvent générer quelques biais propres aux outils d'évaluation utilisés (effet d'attente miroir animateur – participants, effet de l'exemple illustratif sur les participants – ouverture/canalisation/blocage). Ainsi les réponses à cette évaluation restent très orientées : nous avons pu observer que les participants semblent s'être projetés surtout en fonction de leur activité et avoir cherché des exemples parmi ce qu'ils connaissaient déjà. En même temps, compte tenu de la diversité d'activité des participants, cela a également mis en relief la multidisciplinarité des scénagrammes et l'étendue applicative de notre approche.

Dans la suite du projet PRIM, nous allons organiser des ateliers et des séminaires avec des professionnels de santé et des professionnels de l'éducation afin de définir les besoins à couvrir dans les prochaines versions de ScenaProd. Il faudra pour cela présenter beaucoup plus d'exemples de scénagrammes pour éviter de bloquer l'imagination des participants dans une seule voie.